\documentclass[fleqn,10pt]{wlscirep}
\usepackage[utf8]{inputenc}
\usepackage[T1]{fontenc}
\usepackage{subfigure}
\title{Nonlinear Ion-Acoustic Waves with Landau Damping in Non-Maxwellian Space Plasmas}

\author[1,2]{Hadia Mushtaq}
\author[3,1*]{Kuldeep Singh}
\author[2]{Sadia Zaheer}
\author[1,3,4]{Ioannis Kourakis}
\affil[1] { Department of Mathematics, Khalifa University of Science and Technology, Abu Dhabi, UAE}
\affil[2] { Department of Physics, Forman Christian College (A Chartered University), Lahore 54600,  Pakistan}
\affil[3] { Space and Planetary Science Centre, Khalifa University, Abu Dhabi, UAE}

\affil[4] { Hellenic Space Center, Leoforos Kifissias 178, Chalandri, GR-15231 Athens, Greece}

\affil[*] {Corresponding author; email: singh.kdeep07@gmail.com ; kuldeep.singh@ku.ac.ae}

\begin{abstract}
The dynamics of nonlinear ion-acoustic solitary waves in the presence of kinetic (Landau type) damping have been
investigated in a collisionless, non-magnetized electron-ion plasma. A cold ion fluid model, coupled to a Vlasov-type kinetic
equation for the electron dynamics, has been adopted as a starting point. The electron population was assumed to be in
a kappa-distributed state, in account of the non-Maxwellian behavior of energetic (suprathermal) electrons often observed
in Space. A multiscale perturbation technique has led to an evolution equation for the electrostatic potential, in the form of
a modified Korteweg–de Vries (KdV) equation, incorporating a non-local term accounting for Landau damping (associated with the
electron statistics). Exact analytical solutions have been obtained, representing solitary waves undergoing amplitude decay
over time.

The combined effect of Landau damping and non-Maxwellian electron statistics (via the kappa parameter) on the characteristics
of IASWs has been examined. Numerical integration of the evolution equation has been undertaken, to elucidate the importance of kinetic Landau damping on a shock-shaped initial condition.

The results of this investigation aim to improve our understanding of the dynamics of nonlinear electrostatic waves under the influence of Landau damping in various space plasma environments.
\end{abstract}


\begin{document}

\flushbottom
\maketitle
\thispagestyle{empty}

%\noindent Please note: Abbreviations should be introduced at the first mention in the main text – no abbreviations lists. Suggested structure of main text (not enforced) is provided below.

\section{Introduction \label{sec:1} }

Energy localization in dispersive media is known to lead to the formation of stationary profile states (localized modes) with remarkable stability when a balance between dispersion and nonlinearity is attained.
such localized structures come in various forms, including solitary waves, shocks, vortices, double layers and periodic nonlinear waves, among a wide variety of possible nonlinear states.
In the last decades, investigating the characteristics of nonlinear structures in diverse physical contexts has emerged as a forefront area of research. As nonlinear waves propagate, wave steepening due to the intrinsic medium nonlinearity may be compensated either by dispersive effects (giving rise to solitary waves \cite{dauxois}  or by intrinsic dissipation in the medium (leading to shock wave formation).

Solitary waves are often modeled as solitons, i.e. exact solutions of integrable nonlinear partial differential equations (PDEs) \cite{dauxois,infeld}. The Korteweg de-Vries (KdV) equation, a typical representative and the first historical paradigm of such a nonlinear PDE, was first derived for electron-ion plasma by Washimi and Taniuti \cite{washimi66} in their effort to model ion-acoustic solitary waves, thus inaugurating a compelling path for further investigation. The early studies by Taniuti and coworkers \cite{taniuti, washimi66} have subsequently been adopted by the plasma physics community (see e.g. \cite{Ridip})  as an efficient means of describing nonlinear structures, wave-particle interactions, instabilities, and related nonlinear effects, as they occur in various plasma environments, in Space and the laboratory.
Graham et al.\cite{graham16} reported the occurrence of electrostatic solitary waves and field-aligned electrostatic waves near Earth’s magnetopause and in the magnetosheath by using the cluster spacecraft data.
Kakad et al.\cite{kakad22} confirmed and validated the observations of electrostatic solitary waves in the Marian magnetosheath by the Langmuir Probe and Waves instrument on the Mars Atmosphere and Volatile EvolutioN (MAVEN) spacecraft in orbit around Mars. The observed bipolar pulses were identified as ion-acoustic solitary wave structures with a speed close to the ion-acoustic speed.
Steffy et al. \cite{steffy22} investigated the occurrence of various types of solitary waves in the terrestrial auroral plasma. Their findings were shown to match observational evidence, highlighting the indispensable role of the coexistence of two-electron populations on the morphology of localized electrostatic structures.

As regards classical plasmas in particular, two approaches are mainly adopted as working horses by theoreticians in their attempt to describe propagating electrostatic oscillations (plasma waves).
Originally a statistical method by its very nature, plasma kinetic theory relies on defining a probability distribution function (pdf) for each plasma constituent (ions, electrons) and then describing its evolution in time via an appropriate (``kinetic") evolution equation. Measurable macroscopic quantities are then expressed as appropriate moments of the pdf. An alternative approach consists of defining each plasma component as an inertial fluid whose bulk properties (density, fluid speed, ...) are described by suitable fluid-dynamical equations. The self-generated electric or magnetic fields thus appear in the fluid momentum equation(s), which are then coupled via Maxwell's equations. As one might expect, and indeed require from first principles, both approaches lead to (mostly) identical results in terms of the dispersion properties of plasma waves. However, since it is essentially a reduction of the many-body problem into a bulk description, fluid-plasma modeling intrinsically ignores microscopic particle motion and the resonant interaction of particles with the bulk plasma oscillation. This latter effect, known as Landau damping, is predicted by kinetic theory -- but not by fluid theory (a known shortcoming of the latter).

Landau damping \cite{ryutov}
 is a captivating -- and perhaps counterintuitive --  phenomenon inherent to charged matter (plasma), manifested to dissipation (damping) arising -- despite the absence of inter-particle collisions -- due to resonant interactions between waves and particles \cite{dawson, landau} \, . The original prediction in Landau's seminar original paper \cite{landau}  of plasma oscillations undergoing Landau damping, over seven decades ago, was followed by several theoretical experimental studies\cite{ryutov} \, .  Landau originally predicted the collisionless damping of electron plasma waves, specifically in the context of the Langmuir waves, a proposition subsequently experimentally verified by Malmberg and Wharton\cite{malmberg} \, .
 As the nonlinear theory for plasma waves had started gaining momentum, in the late 1960s \cite{taniuti, washimi66}, Ott and Sudan\cite{sud69} elaborated a first model for nonlinear ion-acoustic waves by incorporating the effect of Landau damping due to resonances with a thermal electron background. Not long thereafter, many researchers focused on theoretical and experimental studies of nonlinear Landau damping in different plasma systems \cite{barman, nakamura, das} \, . Barman \cite{barman}  studied Landau damping in an unmagnetized dusty negative-ion plasma with complete electron depletion, where the entire electron population resides on the dust (particulates') surface. Nakamura et al. \cite{nakamura} explored the Korteweg–de Vries equation with Landau damping numerically, and compared their results with experimental findings from a purpose-built experiment based on a bi-ion plasma. Das and Bandyopadhyay\cite{das} have explained the nonlinear evolution equation for IASWs including the effect of Landau damping. The system consisted of warm adiabatic ions and non-thermal electrons in a magnetized plasma. In a recent paper, Ur-Rehman \cite{reh19}  presented a linear kinetic-theoretical analysis predicting  Landau damping in the presence of a suprathermal electron population, in fact focusing on a four-component plasma model allegedly applicable to the 67P/Churyumov–Gerasimenko Comet environment.

Focusing on the second founding pillar of our study, highly energetic (suprathermal) charged particles (i.e. electrons or ions exhibiting a long-tailed velocity trend and a stronger velocity component above the thermal speed than could be explained by a Maxwell-Boltzmann distribution) \cite{liu}. This is a common occurrence in Space observations \cite{pierrard}, and also in the laboratory \cite{sarri}. It is now established that this phenomenon is aptly explained through a \emph{kappa} distribution \cite{hasegawa, summers}, that is a parametrized (non-Maxwellian) velocity distribution function \cite{liva,liva1} with a real parameter ($\kappa$), which appears to be more appropriate than a thermal Maxwellian distribution in a wide range of plasma situations.   This distribution was first suggested by Vasyliunas\cite{vasy} to model space plasmas and was later adopted by many authors in various physical contexts.  Thanks to its ability to explain the power-law dependence observed in Space \cite{oka}, the kappa distribution is now recognized as a ubiquitous feature of Space plasmas \cite{livadiotis} \, . Following its successful recognition as a useful tool in analyzing observations in various regions, e.g. in Earth’s magnetosphere \cite{feldman} and in the auroral region \cite{la,mendis}, the kappa distribution function has not only been used in the interpretation of observed spectra but was also adopted as a more accurate alternative to the  Maxwellian assumption, in theoretical modeling \cite{kourakis, hellberg} \, .

This study aims to investigate, from first principles, the impact of non-Maxwellian electron statistics on the propagation characteristics of ion-acoustic (electrostatic) solitary waves subject to Landau damping. Our ambition is to gain insight into the dynamical mechanisms affecting propagating nonlinear structures (waves) in plasmas and, in particular, to elucidate the role of resonant wave-particle interactions on the formation and propagation of electrostatic solitary waves in Space plasma.

This manuscript is structured as follows. Following this introductory section, an analytical model is laid out in Section \ref{sec:2}, and is adopted as a starting point in our analysis. Adopting the reductive perturbation method, a Korteweg-de Vries (KdV) type equation is then obtained for the evolution of the electrostatic potential, incorporating an additional term in account of Landau damping. Based on that equation, the Landau damping rate for solitary waves in the presence of suprathermal electrons is then computed in the following Section \ref{sec:3} \, .
An analytical solution for the electrostatic potential
% and the other plasma state variables
is derived in Section \ref{sec:4} \, . The parametric analysis follows, in Section \ref{sec:5}, elucidating how the damping mechanisms are affected by the value of the electron kappa  (kappa). The numerical solution of the KdV equation with Landau damping is elaborated in section \ref{sec:7} \, . Finally, our results are summarized in the concluding Section \ref{sec:6} \, .

\section{Analytical model \label{sec:2} }

We have considered an unmagnetized collisionless electron-ion plasma. We shall adopt an infinite one-dimensional (1D) geometry, for simplicity, and will assume the ion temperature to be negligible (cold ion approximation, i.e. $T_{i}=0$),
while the electron temperature $T_{e}$ is finite.
The ion dynamics can be described by the fluid-dynamical equations:
\begin{eqnarray}
\frac{\partial \tilde n_i}{\partial \tilde t} + \frac{\partial (\tilde n_i \tilde u_i)}{\partial \tilde x} = 0 ,\\
\frac{\partial \tilde u_i}{\partial \tilde t} + \tilde u_i \frac{\partial \tilde u_i}{\partial \tilde{x}}=- \frac{e}{m_i}\frac{\partial \tilde{\phi}}{\partial\tilde{ x}}, \\
\frac{\partial^2 \tilde{\phi}}{\partial \tilde{x}^2}=4\pi e(\tilde n_{e} - \tilde n_i) \, . \label{fluid3}
\end{eqnarray}
At equilibrium, the charge neutrality condition in Poisson's equation (\ref{fluid3}) imposes $\tilde n_{e, 0}= \tilde n_{i, 0}= \tilde n_{0}$ (where the subscript `0' denotes the equilibrium values).
The plasma state is thus described by means of four state variables (all functions of space $\tilde x$ and time $\tilde t$): $\tilde n_i$ is the ion number density, $\tilde u_i$ is the ion fluid speed, $\tilde n_{e}$ is the electron number density and $\tilde \phi$ denotes the electrostatic potential. The remaining symbols bear their usual meaning, i.e. $e$ is the (absolute) electron charge and $m_i$ is the ion mass.
The tilde in the above equations (which were here expressed in Gaussian, i.e. CGS units) has been adopted to distinguish the physical variables adopted above (that carry their usual dimensions) from their dimensionless counterparts that will be adopted later in this paper.

The above system is not closed, as the electron number density $\tilde n_{e}$ has not yet been prescribed. Contrary to a commonly adopted approximation, where $\tilde n_{e}$ is assumed to be given by a prescribed equilibrium function  (i.e. neglecting electron inertia), here will shall assume the electron number density to be given by
given by
\begin{equation}
\label{eq:8}
\tilde n_{e}=\int_{-\infty}^{\infty} \tilde f d\tilde v.
\end{equation}
in terms of the (normalized) electron distribution function $f(\tilde x, \tilde v, \tilde t)$.
(where $v$ here refers to the microscopic velocity of the electrons). \color{black}
The evolution of the electron distribution function is described by a kinetic equation in the form of the Vlasov equation \cite{vla68,sin11,Balescu}
\begin{equation}
\frac{\partial \tilde{f}}{\partial \tilde{t}}+\tilde{v} \frac{\partial {\tilde{f}}}{\partial \tilde{x}}+\frac{e}{m_e}\frac{\partial \tilde{\phi}}{\partial \tilde{x}}\frac{\partial {\tilde{f}}}{\partial \tilde{v}}=0 \, .
\end{equation}

For the sake of ease in algebraic calculation, we shall now rescale the quantities entering the above
%fluid-Vlasov
system of equations, as follows
\[u = \frac{\tilde{u}}{c_0}, \qquad  n = \frac{\tilde n_i}{n_0}, \qquad   n_e = \frac{\tilde n_e}{n_{0}}, \color{black} \qquad  f = \frac{c_{e} \tilde{f}}{n_{0}}, \qquad  v = \frac{\tilde{v}}{c_e}, \qquad \phi = \frac{e\tilde{\phi}}{k_B T_e},  \qquad
x=\frac{\tilde{x}}{L}, \qquad  t=\frac{c_0 \, \tilde{t}}{L} \, ,\]
with
\[ c_0=\left(\frac{K_B T_e}{m_i}\right)^{1/2} \qquad  {\rm and} \qquad c_{e}=\left(\frac{K_B T_e}{m_e}\right)^{1/2} \, , \]	
where $K_B$ denotes Boltzmann's constant.
The new dimensionless variables $n$, $u$, $n_{e}$, and $\phi$ denote the ion number density, the ion fluid speed, the electron number density, and the electrostatic potential, respectively.  Notice that the ion fluid speed has been scaled by $c_0$ (which is essentially the plasma sound speed in electron-ion plasma) while the microscopic electron speed has been scaled by the electron thermal speed $c_{e}$. Space and time have respectively been scaled over $L$ and $L/c_0$, where $L$ is a characteristic length (left arbitrary, at this stage).

Henceforth, all algebraic expressions in this article will be dimensionless quantities, i.e. real numbers.
The rescaled (dimensionless) fluid-dynamical system of equations now takes the form 		
\begin{eqnarray}
\frac{\partial n}{\partial t}+\frac{\partial (nu)}{\partial x}=0\label{eq;5},\\
\frac{\partial u}{\partial t}+u \frac{\partial u}{\partial x}=-\frac{\partial \phi}{\partial x}\label{eq:6},\\
\frac{\lambda_{D}^{2}}{L^{2}}\frac{\partial^2 \phi}{\partial x^2}=n_e - n. \label{eq:7}
\end{eqnarray}
Notice that the equilibrium value of both ion and electron number densities now turns out to be equal to unity.
Note the natural appearance of the Debye length $\lambda_{D}= \left( \frac{K_B T_{e} }{4 \pi n_{0} e^{2}}\right)^{1/2}$ through the ratio ${\lambda_{D}^{2}}/{L^{2}}$, that physically represents the strength of wave dispersion due to deviation from charge neutrality (as described by Poisson's equation).
The  (dimensionless)  Vlasov equation now reads\cite{vla68,sin11,Balescu}:		
\begin{equation}
\left(\frac{m_e}{m_i}\right)^{1/2} \frac{\partial f}{\partial t}+v \frac{\partial f}{\partial x}+\frac{\partial \phi}{\partial x}\frac{\partial f}{\partial v}=0 \, ,
\end{equation}
i.e.
\begin{equation}
\label{eq:a}
\delta\frac{\partial f}{\partial t}+v \frac{\partial f}{\partial x}+\frac{\partial \phi}{\partial x}\frac{\partial f}{\partial v}=0 \, ,
\end{equation}
Notice that the quantity
$\left({m_e}/{m_i}\right)^{1/2}$ ($= \delta$)
representing the (finite) electron inertia will henceforth appear as the signature of the kinetic electron dynamics (to be associated with  Landau damping) in the algebraic expressions to follow.

Following the procedure introduced by Ott and Sudan\cite{sud69}, we
shall assume that
\begin{equation}
\label{eq:a1}
\delta = \left(\frac{m_e}{m_i}\right)^{1/2} \approx \gamma_{1} \epsilon, \qquad  \frac{\tilde n_i - n_0}{n_0} = \frac{\tilde n_i}{n_0} - 1= n - 1  \approx  \gamma_2 \epsilon \qquad  {\rm and} \qquad \color{black}  \qquad  \frac{\lambda_{D}^{2}}{L^{2}}  \approx  \gamma_3 \epsilon \, ,
\end{equation}
where $\epsilon \ll 1$ is a small (real-valued) parameter.
Physically speaking, the (three) parameters thus introduced are related to: kinetic electron evolution -- to be later linked to Landau damping -- for $\gamma_1$, nonlinearity -- i.e. to the strength of the deviation from equilibrium -- for $\gamma_2$ and to dispersion -- as discussed above, in Poisson's equation-- for $\gamma_3$, here all assumed to be small and, in fact, comparable in order of magnitude.
%\begin{equation}
%\frac{n}{n_0}=\gamma_2 \epsilon,
%\end{equation}
%\begin{equation}	
%\frac{\lambda_{D}^{2}}{L^{2}}=\gamma_3 \epsilon
%\end{equation}
Eq. \eqref{eq:a} may now be rewritten as
\begin{equation}
\label{eq:14}
\gamma_{1} \epsilon\frac{\partial f}{\partial t}+v \frac{\partial f}{\partial x}+\frac{\partial \phi}{\partial x}\frac{\partial f}{\partial v}=0.
\end{equation}

The reductive perturbation method\cite{taniuti,sud69} will now be adopted, in view of the derivation of an evolution equation describing the dynamics of small amplitude nonlinear electrostatic excitations. Anticipating stationary-profile localized solutions moving at speed $V$, we shall first introduce stretched space and time coordinates in the form
\begin{equation}
\label{Eq:15}
\xi=\epsilon^{\frac{1}{2}}(x-V t) \, , \qquad \tau=\epsilon^{\frac{3}{2}} t 	\, ,
\end{equation}
where $V$ will be determined later by algebraic requirements.
Subsequently, all dependent variables will be expanded in powers of the (nonlinearity related) coefficient, as
\begin{eqnarray}
n &=& 1+\gamma_{2} \, \epsilon \, n^{(1)}+ \gamma_{2}^{2} \, \epsilon^{2} \, n^{(2)}+...\\
u &=& \gamma_{2} \, \epsilon u^{(1)}+\gamma_{2}^{2} \, \epsilon^{2} u^{(2)}+...\\
\phi &=& \gamma_{2} \, \epsilon \, \phi^{(1)}+\gamma_{2}^{2} \epsilon^{2} \phi^{(2)}+...\\
n_{e} &=& 1+\gamma_{2} \, \epsilon n_{e}^{(1)}\, +\gamma_{2}^{2} \epsilon^{2} n_{e}^{(2)}+...\\
f &=& f^{(0)}+\gamma_{2} \, \epsilon \, f^{(1)}+\gamma_{2}^{2} \epsilon^{2} f^{(2)}+...
\end{eqnarray}
$f^{(0)}$ is the equilibrium distribution function. We shall adopt the kappa distribution here\cite{hellberg}, in the form
\begin{equation}
\label{eq:21}
f^{(0)}=\frac{1}{\sqrt{\pi}\theta}\frac{\Gamma(\kappa+1)}{\kappa^{\frac{3}{2}}\Gamma(\kappa-\frac{1}{2})}\left(1+\frac{v^{2}}{\kappa \theta^{2}}\right)^{-\kappa}, \qquad
\theta=\left( \frac{2\kappa-3}{\kappa}\right)^{\frac{1}{2}} \, .
\end{equation}
Note the Maxwell-Boltzmann equilibrium function
\begin{equation}
\label{Maxwellian}
f_{MB}^{(0)}=\frac{1}{\sqrt{2\pi}} \exp\left(-\frac{v^{2}}{2}\right).
\end{equation}
is recovered in the limit of infinite $\kappa$ (value).

\section{Evolution equation for stationary-profile nonlinear structures}

We may now proceed by substituting the above polynomial expansions (in powers of $\epsilon$) into the evolution equations and then collecting and analyzing the terms arising in various orders. The lengthy algebraic procedure is tedious but perfectly straightforward. Details can be found in the supplementary information, \color{black}, and are thus omitted here.

The analytical process leads to a set of relations connecting the leading order corrections (disturbances) to the equilibrium values, i.e.
\begin{equation}
\label{eq:a2}
n^{(1)} = \frac{1}{V} u^{(1)} = \frac{1}{V^2}\phi^{(1)}  \, ,
\end{equation}
where the phase speed $V$ is found by algebraic compatibility requirements to be given by
\begin{equation}
V = \frac{1}{\sqrt{a_1}} = \left(\frac{\kappa-\frac{3}{2}}{\kappa-\frac{1}{2}}\right)^{1/2} \, .
\end{equation}
It is important to notice that the latter expression is essentially the ion-acoustic (sound) speed in e-i plasma, in the presence of suprathermal electrons, as derived and analyzed in earlier works on electrostatic waves in kappa-distributed plasmas\cite{kourakis} \, .

Moving on to higher orders in $\epsilon$, a lengthy algebraic procedure (outlined in the appendix, i.e. as Supplemental Material to this article)
leads to an evolution equation in the form:
\begin{equation}
  \label{eq:KdV}
\frac{\partial n}{\partial\tau}+A \, n \, \frac{\partial n}{\partial \xi}+B\frac{\partial^{3} n}{\partial\xi^{3}}+ C \, P\int_{-\infty}^{+\infty}\left( \frac{\partial n} {\partial \xi^{'}}\right) \frac{d\xi^{'}}{\xi-\xi^{'}}=0 \, ,
\end{equation}
that is, essentially a Korteweg - de Vries (KdV) type\cite{dauxois}  PDE with an additional term (see the last term in the LHS), in account of the kinetic electron dynamics.   Note that the dependent variable in the latter equation is essentially the leading (first) order density disturbance $n^{(1)}$, that will henceforth be denoted a $n$ (i.e. dropping the superscript) throughout what follows.
The (real) coefficients appearing in (\ref{eq:KdV}) are all functions of $\kappa$; these are:

--- the nonlinearity coefficient $A$, given by
\begin{equation}
    A = \frac{\gamma_{2}}{2}\left(3V-a_{3}V^{5} \right)=\frac{\gamma_{2}}{2}\left[3\left(\frac{\kappa-\frac{3}{2}}{\kappa-\frac{1}{2}}\right)^{\frac{1}{2}}-\frac{(\kappa+\frac{1}{2})(\kappa-\frac{3}{2})^{\frac{1}{2}}}{(\kappa-\frac{1}{2})^{\frac{3}{2}}} \right] \, ,
\end{equation}

--- the dispersion coefficient $B$, given by
 \begin{equation}B=\frac{\gamma_{3}V^{3}}{2}=\frac{\gamma_{3}}{2}\left(\frac{\kappa-\frac{3}{2}}{\kappa-\frac{1}{2}}\right)^{\frac{3}{2}},\,  \end{equation}
and

 --- a kinetic ``damping" coefficient $C$, given by
\begin{equation}
C = \frac{V^{3}a_{2}}{2}=\frac{\gamma_{1}}{2\sqrt{2\pi}} \left[\frac{\Gamma(\kappa+1)(\kappa-\frac{3}{2})^{\frac{1}{2}}}{(\kappa-\frac{1}{2})^{2} \, \Gamma(\kappa-\frac{1}{2})}\right] \, .\end{equation}
The last term in the LHS of the KdV equation above is associated with energy dissipation in the form of Landau damping, and in fact, breaks the integrability of the above PDE (known to be integrable if $C=0$). This term is therefore expected to result in damped (i.e. decaying amplitude) solutions, as will be shown in the following.

  In the latter formulas, we have used the definitions:
\begin{equation}
a_{1} = \frac{\kappa-\frac{1}{2}}{\kappa-\frac{3}{2}}, \qquad a_{2} = \frac{\gamma_{1}}{\sqrt{2\pi}} \left[\frac{\Gamma(\kappa+1)}{(\kappa-\frac{1}{2})^{\frac{1}{2}}(\kappa-\frac{3}{2}) \, \Gamma(\kappa-\frac{1}{2})}\right] \qquad {\rm and} \qquad a_{3} =
\frac{\kappa^{2}-\frac{1}{4}}{\left(\kappa-\frac{3}{2}\right)^{2}} \, ,
\end{equation}
where $\Gamma$ denotes the Gamma function
%that is a generalization of factorial function to real and complex numbers. It is often used in probability and statistics, as it shows up in the normalizing constants of probability distributions
\cite{abramowitz} \, .

Some limiting cases may be worth discussing at this point.
First of all,
in the limit $\kappa \rightarrow \infty$, one recovers
\(
A=\gamma_{2}, \quad B= {\gamma_{3}}/{2} \quad {\rm and} \quad C= {\gamma_{1}}/(2\sqrt{2\pi})
\),
i.e.
the results match exactly with the expressions obtained by Sudan \& Ott \cite{sud69} (for the Maxwellian electron case).
If, furthermore, one neglects the Landau damping effect by taking $C=0$ (or $\gamma_1=0$), thus neglecting the electron inertia, and then sets  $\gamma_2=\gamma_3=1$, on recovers the standard KdV equation with $A=1$ and $B=1/2$ for Maxwellian e-i plasma, as originally obtained by Washimi \& Taniuti\cite{washimi66} \, .
On the other hand, if one retains finite $\kappa$ values (in the account of suprathermal electrons), but neglects electron inertia (i.e. upon setting $\gamma_1 =0$), one finds the known expressions for the KdV coefficients associated with electrostatic solitary waves in kappa-distributed plasmas\cite{kourakis} \, .

\section{Landau damping Rate for ion-Acoustic Waves in the presence of suprathermal particles
\label{sec:3}}

To explore electrostatic wave damping, let us set $A=B=0$ for a minute in the evolution equation (\ref{eq:KdV}), reducing it to
\begin{equation}
\label{eq:48}
 \frac{\partial n}{\partial\tau}+  C \, P \, \int_{-\infty}^{+\infty}\left( \frac{\partial n} {\partial \xi^{'}}\right) \frac{d\xi^{'}}{\xi-\xi^{'}}=0 \, ,
\end{equation}
where $P$ denotes the Cauchy principal value. Now, we proceed by using the convolution theorem on the second term -- see Eq.(\ref{A2}) in the Appendix (supplemental material), then taking the Fourier transform of Eq. (\ref{eq:48}) and finally using the Fourier formula $F \left[ P\left(\frac{1}{\xi}\right)\right]= -i\, \pi \, |k|$. Finally,  linearizing the Fourier transformation of $f$ -- cf. Eq. \ref{A1} in the Appendix (supplemental material) -- one obtains:
%
% \[- i \omega+ C \pi|k|=0\]
%
\begin{equation}
\label{eq:49}
    \omega= i \, \pi \, C \, |k| \, ,
\end{equation}
where $\omega$ is the angular frequency and $k$ is the wavenumber associated with the electrostatic wave.
The frequency thus becomes imaginary, accounting for a normalized linear damping decrement (rate) given by
\begin{equation}
\label{eq:50}
    |\gamma|= \pi \, C \, .
\end{equation}

\begin{figure}[ht]
\centering
\includegraphics[width=3in]{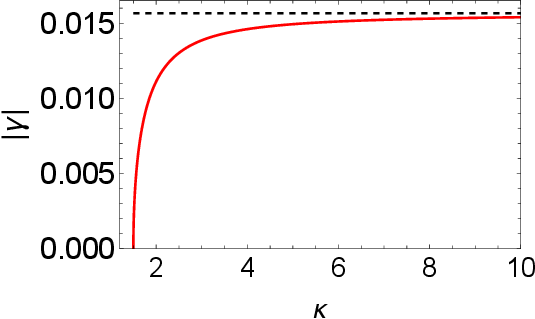}
\caption{Variation of the linear Landau damping rate ($|\gamma|$) versus the parameter $\kappa$.}
\label{f1}
\end{figure}

The linear Landau damping decrement, associated with the coefficient $C$, has been depicted in Fig. \ref{f1} versus kappa ($\kappa$). One easily sees that, as the value of $\kappa$ decreases (i.e., on account of a stronger deviation from the thermal equilibrium state), the linear Landau damping rate decreases too. It appears that the non-Maxwellian character of the electron statistics thus impedes the kinetic damping mechanism.  Note that,  for $\kappa \rightarrow \infty $, the asymptotic value of  $\gamma$ is $\gamma=\frac{\gamma_1}{2}\sqrt{\frac{\pi}{2}} \simeq 0.01566$ for fixed $\gamma_{1}\approx \left({m_e}/{m_i}\right)^{1/2} \approx 0.025$. In Fig.\ref{f1}, the asymptotic value is represented by the black horizontal dashed line attained for large kappa (values).

 \begin{figure}[ht!]
\centering
\includegraphics[width=3in]{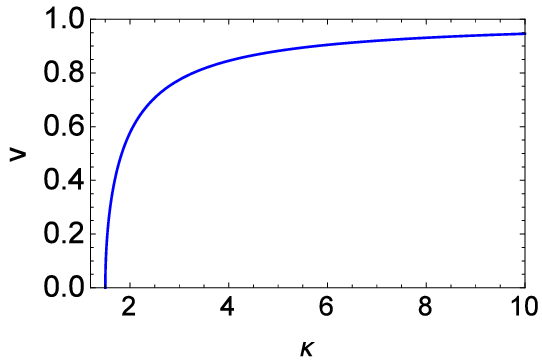}
\caption{Variation of the phase speed ($V$) vs kappa ($\kappa$).}
\label{f2}
\end{figure}

\begin{figure}[ht!]
\centering
\includegraphics[width=3in]{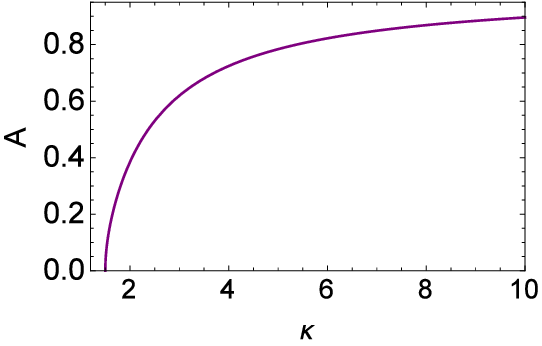}
\caption{Plot of the nonlinearity coefficient ($A$) vs kappa  ($\kappa$).}
\label{f3}
\end{figure}

\begin{figure}[ht!]
\centering
\includegraphics[width=3in]{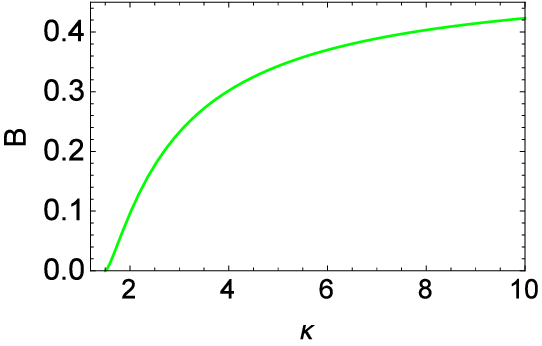}
\caption{Plot of the dispersion coefficient ($B$) vs kappa  ($\kappa$).}
\label{f4}
\end{figure}
\begin{figure}[ht!]
\centering
\includegraphics[width=3in]{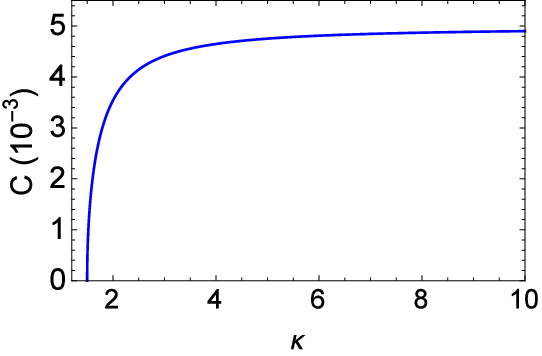}
\caption{Plot of the Landau damping coefficient ($C$) vs kappa  ($\kappa$).}
\label{f5}
 \end{figure}

\begin{figure}[ht!]
\centering
\includegraphics[width=3in]{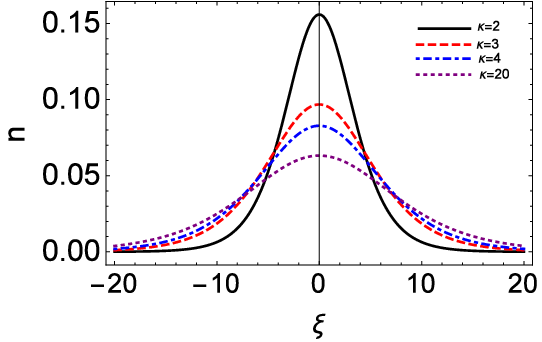}
\caption{The solitary wave (pulse) form ($n$) vs moving reference coordinate  ($\xi$) setting $\tau = 0$. This plot is based on Eq. (\ref{eq:51}). }
\label{f6}
 \end{figure}

\begin{figure}[ht]
\centering
\includegraphics[width=3in]{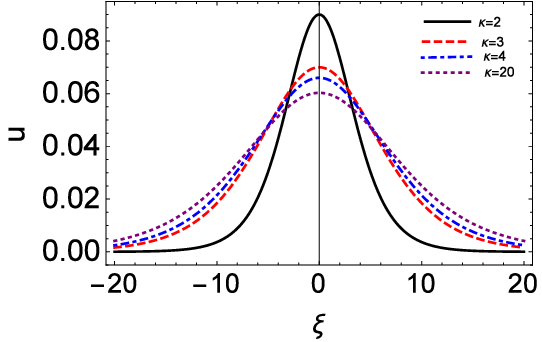}
\caption{The ion fluid speed (u) vs moving reference coordinate  ($\xi$) setting $\tau = 0$. This plot is based on Eq. \eqref{eq:a2} i.e. $u \simeq V n$ (to leading order in $\epsilon$). } \, .
\label{f7}
 \end{figure}

\begin{figure}[ht!]
\centering
\includegraphics[width=3in]{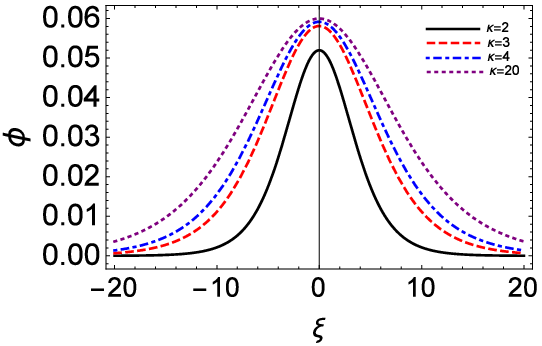}
\caption{Variation of the electrostatic potential excitation ($\phi$) vs moving reference coordinate  ($\xi$) setting $\tau = 0$. This plot is based on Eq. \eqref{eq:a2} i.e. $\phi \simeq V^2 n$ (to leading order in $\epsilon$).  }
\label{f8}
\end{figure}

\begin{figure}[ht!]
\centering
\includegraphics[width=3in]{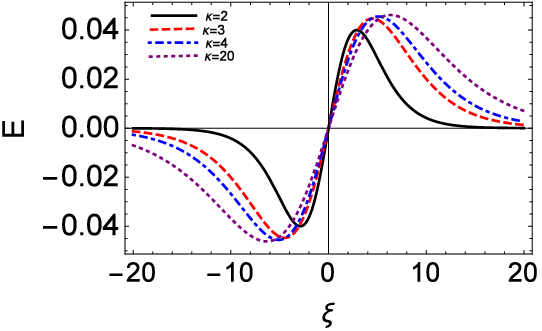}
\caption{The electric field (E) (bipolar) waveform vs moving reference coordinate  ($\xi$) setting $\tau = 0$.}
\label{f9}
\end{figure}

\begin{figure}[ht!]
\centering
\includegraphics[width=3in]{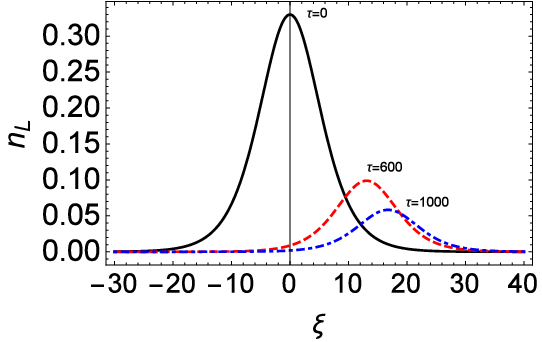}
\caption{Time evolution of the solitary wave profile ($n_{L}$)  under the influence of Landau damping for fixed $\kappa =4$.}
\label{f10}
\end{figure}

\begin{figure}[ht!]
\centering
\includegraphics[width=3in]{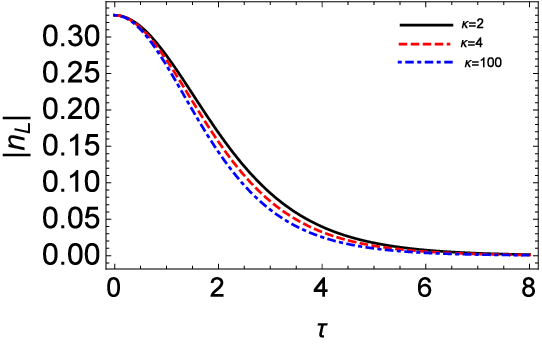}
\caption{The variation of the peak amplitude ($|n_{L}|$) of the soliton pulse vs time for different kappa ($\kappa$) values.}
\label{f11}
\end{figure}

\begin{figure}
\centering
\subfigure[]{\includegraphics[width=2.5in]{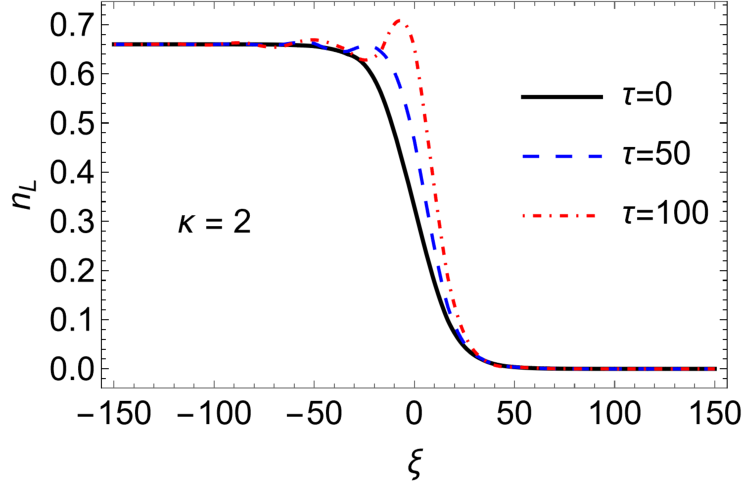}}\hspace{0.5in}
\subfigure[]{\includegraphics[width=2.5in]{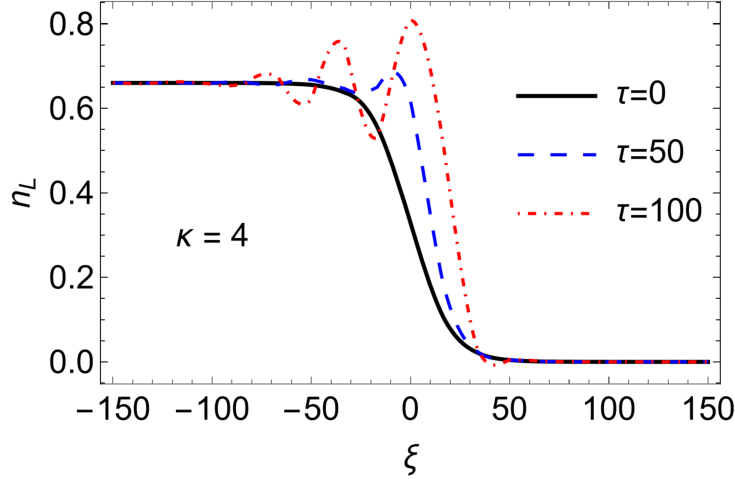}}
\subfigure[]{\includegraphics[width=2.5in]{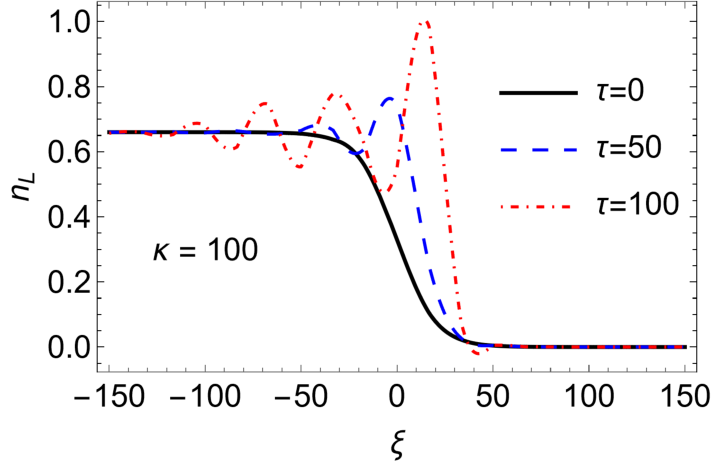}}
\caption{Numerical solution of KdV equation with Landau damping in the time domain for different values of $\kappa$. Here $\rho=0.05$ and $\Theta=0.33$ are fixed.}\label{f12}
\end{figure}

\section{Analytical solution of the KdV equation under the effect of Landau damping
\label{sec:4} }

In the absence of damping, i.e. for $C=0$, the resulting KdV Eq. \eqref{eq:KdV} possesses a solitary wave solution in the form
\begin{equation}
\label{eq:51}
 n=N_{0} \, \mathrm{sech}^{2}\left(\frac{\xi-U_{0}\tau}{W}\right) \, ,
\end{equation}
Here, $N_0$ is a constant, representing the soliton (pulse) amplitude (note that this will be a constant for $C=0$ -only - i.e. in the absence of damping).
This is a monoparametric family of solutions, dictating the maximum pulse amplitude and the pulse width to be given by to be $N_0 = {3 U_{0}}/{A}$ and  $W=\sqrt{\frac{4B}{U_{0}}}$ respectively, where $U_{0}$ represents the (arbitrary) pulse speed in the moving reference frame\cite{shukla,shukla2} \, .

Now,
we may attempt to calculate the effect of Landau damping on the solitary wave solution of the KdV equation, by adopting the procedure proposed by Barman \& Misra\cite{barman} \, . In view of performing a perturbation analysis, we shall assume that $C (\gg \epsilon)$ is smaller than the other parameters in order of magnitude, viz. $A \approx B\gg C$ ($\gg \epsilon$) \cite{barman} \, .
Let us  introduce a new space coordinate in a moving frame, that moves at the wave's (time-varying) speed:
\begin{equation}
\label{eq:52}
z = \frac{1}{W}\left( \xi-\frac{A}{3} \int_{0}^{\tau}N(\tau^{'}; C) \, d\tau^{'}\right) \, .
\end{equation}
Here, $N=N\left(\tau; C\right)$ denotes the soliton amplitude, which is a slowly decreasing function of time, under the effect of damping, i.e. for $C\ne 0$; cf. $N_0 = N\left(\tau; C = 0\right)$, which appears in (\ref{eq:51}) above. Likewise, the density (disturbance) will be denoted by $n=n\left(\tau; C\right)$.

  After this transformation, Eq. (\ref{eq:KdV}) becomes
\begin{equation}
\label{eq:53}
\frac{\partial n}{\partial \tau}+\frac{C}{W}P\int_{-\infty}^{\infty}\frac{\partial n}{\partial z^{'}}\frac{d z^{'}}{z-z^{'}}-\left(\frac{N A}{3 W}-\frac{z}{2N} \frac{dN}{d\tau}  \right)\frac{\partial n}{\partial z}+ \frac{A}{W} n \frac{\partial n}{\partial z}+ \frac{B}{W^{3}}\frac{\partial^{3}n}{\partial z^{3}}=0 \, ,
\end{equation}
where we used $\frac{\partial n}{\partial z^{'}}=\frac{\partial n}{\partial z}$ at $z=z^{'}$.
%as shown in \cite{misra} \, .
To obtain an analytical solution of Eq. \eqref{eq:53}, we have followed the procedure adopted by Ott \& Sudan\cite{sud69} \, . Furthermore, we  generalize the multiple time scale technique by exploiting the smallness of $C$ \cite{barman,yashika}: we consider a solution in the form
 \begin{equation}
  \label{eq:54}
n(z,\tau)=n^{(0)}+ C \, n^{(1)}+ C^{2} \, n^{(2)}+C^{3} \, n^{(3)}+.....
\end{equation}
An algebraic calculation leads to an analytical solution of Eq. \eqref{eq:KdV} in the form
\begin{equation}
\label{eq:55}
n(\tau) = N_0 {\left( 1+\frac{\tau}{\tau_{0}}\right)}^{-2} \mathrm{Sech}^{2} \left(\frac{\xi-\frac{A}{3}\int_{0}^{\tau}N(\tau^{'})d \tau^{'}}{W}\right) + {\cal O}(C) \, ,
\end{equation}
where
%$\tau_0$ is
\begin{equation}
\label{eq:56}
\tau_{0}= \frac{1.37}{C}\sqrt{\frac{3 B}{A N_0}} \, .
\end{equation}
Note that the latter expression has been obtained at leading order in (i.e. retaining on the zeroth power of) $C$.
This represents a damped soliton (pulse) form,
%subject to Landau damping,
whose amplitude $N(\tau) = N_0\left( 1+\frac{\tau}{\tau_{0}}\right)^{-2}$ decays in time, while also slowing down: note that its instantaneous speed, given by
\begin{equation}
\label{eq:57}
U(\tau) =\frac{1}{\tau}\frac{N_{0}A}{3}\int_{0}^{\tau}\left( 1+\frac{\tau^{'}}{\tau_{0}}\right)^{-2} d \tau^{'} .\,
\end{equation}
is a decreasing function of time.
Note that, upon setting $\tau =0$ in Eq. (\ref{eq:55}), one recovers precisely Eq. (\ref{eq:51}), as expected.

\section{Parametric investigation
\label{sec:5} }

We may now investigate the parametric influence of various plasma (configurational) parameters on the properties of electrostatic solitary waves, as these are subject to kinetic (Landau) damping. The variation of various quantities will be discussed, based on Figs. (\ref{f2})-(\ref{f11}).
To begin with, the phase velocity $V$ increases with kappa ($\kappa$), as shown in Fig. \ref{f2} \, . Stronger deviation of the electron statistics from the Maxwell-Boltzmann equilibrium profile (i.e., for smaller $\kappa$) will therefore enable solitary waves to propagate at a slower speed (than in the Maxwellian case). Recalling that $V$ is essentially the true sound speed in our case (i.e. for an e-i plasma with kappa distributed electrons), we realize that this conclusion simply recovers an earlier result\cite{kourakis}, which is associated with the expected variation of the charge screening (Debye) radius to $\kappa$ \cite{bryant, mc} \, .

In the following, ad hoc values of the real parameters (i.e., $\gamma_2 = \gamma_3 = 1$) have been adopted while producing Figs.  \ref{f3} to \ref{f12} \, .

Figures \ref{f3}, \ref{f4} and \ref{f5} display the variation of the nonlinearity (A), dispersion (B), and kinetic damping (C) coefficient(s)  with kappa ($\kappa$). All of these are increasing functions of $\kappa$. Note that only positive polarity structures will be formed as the nonlinearity coefficient $A$ associated with the soliton amplitude is a positive quantity for all values of $\kappa$.

Fig. \ref{f6} depicts the number density, as it is prescribed by the analytical solution of the KdV equation, at time zero, i.e. ignoring the Landau damping effect: see Eq. \eqref{eq:47} \, . We see that the amplitude of compressive solitary waves increases for lower $\kappa$ (i.e. for stronger deviation from the Maxwellian picture). A similar trend is witnessed in the ion fluid speed (profile) shown in  Fig. \ref{f7} \, . The suprathermal electrons thus seem to lead to stronger compression but to slower ion-acoustic excitations of the ion-fluid properties (in fact, the lower the value of kappa, the stronger the localized compression but also the fluid speed suppression will be).   Note that the latter two figures have been plotted on the moving reference frame, i.e. by setting $\tau = 0$ in the moving coordinate acting as the argument in the respective functions.

The electrostatic potential and the associated electric field excitation predicted for different values of $\kappa$ are shown in Figs. \ref{f8} and \ref{f9}, respectively.
We see that the amplitude of the electrostatic potential profile increases as the values of the kappa increase.  A similar effect can be seen from the electric field profile of IA solitary waves. In these plots, no Landau damping effect is incorporated yet.

 Fig. (\ref{f10}) shows the temporal evolution of IA solitary waves under the influence of Landau damping. Over time, the amplitude of the IASWs tends to decrease due to the Landau damping term. It is also noticed that for higher values of $\kappa$,  the amplitude of the density profile will be shorter as seen in Fig (\ref{f11}).

 \section{Numerical solution of KdV equation with Landau damping \label{sec:7}}

The interplay between nonlinearity and dissipation, in our case attributed to Landau damping, is known from earlier works on qualitatively related studies based on ad hoc damping terms\cite{mal96,sul12,ik12,elk18,singh24} leads to the formation of shock-like formations within the plasma. These shock waves are manifested as propagating fronts, i.e. regions of a sudden transition between different values in the plasma properties.  It is tempting to explore how altering the statistics (i.e. the suprathermal nature) of the electrons might impact these Landau damping-adjusted shock-like patterns.

Employing a finite difference scheme,
%in MATHEMATICA,
we have focused on tracking the temporal evolution of a propagating front (shock-like) excitation subject to Landau damping. In search of a time-dependent numerical solution, we have integrated the KdV equation by adopting a shock-like  step-profile, given by\cite{yashika}
\begin{equation}
n_L(0 \, ,\xi)=\Theta \left[1-{\rm tanh}(\rho \xi)\right] \, , \qquad \xi \in (-\Delta, \Delta)
\end{equation}
where $\Delta$ regulates the dimension of the box, $\Theta$ (an arbitrary constant) is the shock amplitude and $\rho$ is the inverse width of the shock (i.e. this antikink form will be wider, for smaller $\rho$ values). The boundary conditions are $n_L(\tau,-\Delta)=\Theta\left[1+ {\rm tanh}(\rho \Delta)\right]$ and $\phi(\tau, \Delta)=0$. Here, We have assumed $\Delta=150$ for numerical simulation.

Figures \ref{f12} (a, b, c) depict the numerical solutions of the KdV equation incorporating Landau damping across various $\kappa$ values. Analysis of these figures indicates that the initial shock-like compression pulse undergoes amplification and develops an oscillatory tail over time. This phenomenon arises from the increasing dominance of dispersive effects in plasma dynamics, leading to a transition from a monotonic shock to an oscillatory shock profile. Notably, in Figure \ref{f12}(a), at lower $\kappa$ values (i.e., exhibiting high superthermality), dispersive effects exert less influence, thereby preserving the monotonic nature of the shock wave for an extended duration. Conversely, Figures \ref{f12}(b) and (c) illustrate that as $\kappa$ values approach the Maxwellian limit, the shock-like compression pulse amplifies and acquires an oscillatory tail more rapidly. In essence, this implies that monotonic shock structures exhibit greater temporal stability in highly superthermal cases compared to the Maxwellian case. Moreover, the effect of the Landau damping on the shock-like compression pulse undergoes amplification and develops an oscillatory tail over time for highly superthermal cases and the Maxwellian case has been encapsulated in an animation video (see the supplementary video).

\section{Conclusion
\label{sec:6} }

We have examined the combined effect of nonthermal (non-Maxwellian) electron statistics and kinetic (Landau) damping on the propagation characteristics of ion-acoustic (IA) solitary waves in electron-ion plasma.
By employing a multiple-scale (variable stretching) technique, a nonlinear PDE in the form of a Korteweg - de Vries (KdV) equation was obtained, in terms of the leading-order density disturbance. This equation is a modified version of the historic KdV, here enriched by the appearance of a dissipative term, in account of kinetic damping.

Analysis has shown that the solitary wave (pulse-shaped) solution decreases over time, as expected,  due to the addition of the kinetic (Landau) damping term to the evolution equation. The combined parametric influence of kinetic damping and suprathermal electron statistics on the pulse characteristics has been examined.

In a separate line of investigation, We have numerically integrated the nonlinear evolution equation with a step-shaped initial condition (reminiscent of an antikink soliton form). The initial condition was seen to decompose into a series of pulses as time progressed.
Nonetheless, stronger deviation from the Maxwell-Boltzmann distribution for the electron background appears to slow down this instability of the initial condition.

In the limiting case $\kappa \rightarrow \infty$, our results are in good agreement with the Maxwellian case, studied in earlier works.

The findings of this investigation may shed light on the dynamics of nonlinear waves under the influence of Landau damping in various space and astrophysical environments where superthermal electrons are observed.

\section*{Acknowledgment}

Authors KS and IK also acknowledge financial support from Khalifa University’s Space and Planetary Science Center under grant No. KU-SPSC-8474000336. Authors HM, KS, and IK also gratefully acknowledge financial support from Khalifa University of Science and Technology, Abu Dhabi UAE via the (internal funding) project FSU-2021-012/8474000352. IK also acknowledges support from a CIRA (Competitive Internal Research) CIRA-2021-064/8474000412 grant.

\section*{Author contributions}
All authors contributed to the study conceptualization, formal design, and methodology. All authors contributed to the analysis of the results. A first complete draft was written by HM and the manuscript was finalized by KS and IK. All authors have read and approved of the final manuscript.

\section*{Declaration of competing interest}
The authors do not have any kind of conflict of interest.

\section*{Data availability}

All data generated or analysed during this study are included in this published article.

\newpage

\appendix

 \section*{Supplemental Material: Derivation of Eq. (\ref{eq:KdV})}

Using the variable stretching Ansatz Eq. \eqref{Eq:15}, we have:
$\frac{\partial}{\partial x}=\epsilon^{\frac{1}{2}}\frac{\partial}{\partial \xi} \qquad {\rm and} \qquad \frac{\partial}{\partial t}=-V\epsilon^{\frac{1}{2}}\frac{\partial}{\partial\xi}+\epsilon^{\frac{3}{2}}\frac{\partial}{\partial \tau}$ \, .

\subsection*{Leading-order perturbation $\epsilon^{\frac{3}{2}}$:}		
%The lowest rate of $\epsilon$,
Equating the coefficients of $\epsilon^{\frac{3}{2}}$ from Eq. \eqref{eq;5}-\eqref{eq:7} and Eq. \eqref{eq:14}
\begin{eqnarray}
-V \frac{\partial n^{(1)}}{\partial \xi}+ \frac{\partial u^{(1)}}{\partial \xi}=0 \, ,\label{eq:22}\\
-V \frac{\partial u^{(1)}}{\partial \xi}+\frac{\partial \phi^{(1)}}{\partial \xi}=0 \, ,\label{eq:23}\\
n_{e}^{(1)}-n^{(1)}=0 \, ,\label{eq:24}
\end{eqnarray}
and from the Vlasov equation,		
\begin{equation}
\label{eq:25}
v \frac{\partial f^{(1)}}{\partial \xi}+\frac{\partial \phi^{(1)}}{\partial \xi}\frac{\partial f^{(0)}}{\partial v}=0 \, ,
\end{equation}
where $v$ denotes the microscopic velocity of the electrons. After integrating  Eqs. \eqref{eq:22} and  \eqref{eq:23}, we obtain:
\begin{eqnarray}
n^{(1)}=\frac{1}{V^2}\phi^{(1)} \qquad {\rm and} \qquad u^{(1)}=\frac{1}{V}\phi^{(1)} \, .\label{eq:27}
\end{eqnarray}

%Now, Eq. \eqref{eq:25} yields,
%{\color{red}
%\begin{equation} \label{eq26}
%\frac{\partial f^{(1)}}{\partial \xi}=-\frac{\partial f^{(0)}}{\partial v}\frac{\partial \phi^{(1)}}{\partial \xi}+\lambda(\xi,\tau)\delta(v),
%\end{equation}}
%where $\lambda(\xi,\tau)$ is function of $\xi$ and $\tau$ and $\delta(v)$ is Dirac’s delta function\cite{yashika} \, .
To obtain a unique solution of Eq. (\ref{eq:25}), we shall include ad hoc in Eq.
 (\ref{eq:25}) a higher-order term
%$\epsilon^{\frac{7}{2}}\gamma_{1}\epsilon^{2} \frac{\partial f^{(1)}}{\partial \tau}$
originating from the third-order expressions in    $\epsilon$, i.e.\cite{sud69}
\begin{equation}
\label{eq:29}
\gamma_{1}\epsilon^{2} \frac{\partial f^{(1)}_{\epsilon}}{\partial \tau} + v \frac{\partial f^{(1)}_{\epsilon}}{\partial \xi}+\frac{\partial \phi^{(1)}}{\partial \xi}\frac{\partial f^{(0)}_{\epsilon}}{\partial v}=0 \, .
\end{equation} \color{black}
The solutions of the initial value problems \eqref{eq:29} can now be found uniquely, once $f^{(1)}_{\epsilon}$  is  known, by letting $\epsilon \rightarrow 0$ as:
\(f^{(1)}=\lim_{\epsilon\rightarrow 0} f_{\epsilon}^{(1)} \, .\)
Now, taking the Fourier transform of the above equation,
\begin{equation}
f(\omega,k)=\int_{-\infty}^{\infty}\int_{-\infty}^{\infty} f(\xi,\tau)\exp i(k\xi-\omega \tau) \, d\xi \, d\tau \, ,\label{A1}
\end{equation}
we obtain
\begin{equation}
f^{(1)}=\lim_{\epsilon\rightarrow 0}\left(\frac{-k \frac{\partial f^{(0)}}{\partial v}}{k v- \omega \gamma_{1}\epsilon^{2}}\right)\phi^{(1)} \, . \label{eq:31}\end{equation}
Recalling the identity
\begin{equation}
\lim_{\epsilon\rightarrow 0}\left(\frac{1}{k\omega- \omega\gamma_{1}\epsilon^{2}}\right)=P\left(\frac{1}{k v}\right)+ i \, \pi \, \delta(k v) \, ,
\end{equation}		
where $P$ denotes the Cauchy principal value and $\delta$ is Dirac's delta function, we can rewrite Eq. \eqref{eq:31} as:		
\begin{equation}
f^{(1)}=-k\frac{\partial f^{(0)}}{\partial v}\left[P\left(\frac{1}{k v}\right)+i \pi \delta(k v)\right] \phi^{(1)}=-2\frac{\partial f^{(0)}}{\partial v^{2}}\phi^{(1)} \, .
%& =&-\frac{1}{v}\frac{\partial f^{(0)}}{\partial v}[(k v) P (\frac{1}{k v})+\frac{i\pi}{v}\frac{\partial f^{(0)}}{\partial v}(kv)\delta (k v)]\\
%& =& -\frac{1}{v}\frac{\partial f^{(0)}}{\partial v}\phi^{(1)}\\
%& =& -\frac{1}{v} (\frac{\partial f^{(0)}}{\partial v^{2}})\frac{\partial v^{2}}{\partial v}\phi^{(1)}\\
\end{equation}
		
%\begin{equation}   \label{eq:32} f^{(1)}= -2\frac{\partial f^{(0)}}{\partial v^{2}}\phi^{(1)} \, . \end{equation}
From Eqs. \eqref{eq:8} and \eqref{eq:21}, we obtain:
\begin{equation}
n_{e}^{(1)}=\int_{-\infty}^{+\infty} -2 \frac{\partial}{\partial v^{2}}\left[\frac{1}{\sqrt{\pi}\theta}\frac{\Gamma(\kappa+1)}{\kappa^{\frac{3}{2}}\Gamma(\kappa-\frac{1}{2})}\left(1+\frac{v^{2}}{\kappa \theta^{2}}\right)^{-\kappa}\right]\phi^{(1)} dv = \left(\frac{\kappa-\frac{1}{2}}{\kappa-\frac{3}{2}}\right)\phi^{(1)}=a_1 \phi^{(1)} \, .\label{eq:32a}
\end{equation}

%putting this value in Eq. \ref{eq:24}
%\[\frac{\kappa-\frac{1}{2}}{\kappa-\frac{3}{2}}\phi^{(1)}-n^{(1)}=0\]
Combining  Eqs. \eqref{eq:27} and \eqref{eq:32a} into \eqref{eq:24}, we get
\[a_{1}\phi^{(1)}-\frac{1}{V^2}\phi^{(1)}=0 \, . \]
The phase speed is thus prescribed as
\[V=\frac{1}{\sqrt{a_1}} \, , \qquad {\rm where} \qquad  a_1=\left(\frac{\kappa-\frac{1}{2}}{\kappa-\frac{3}{2}}\right).\]

\subsection* {Higher-order perturbation $\epsilon^{\frac{5}{2}}$}
		
%The next higher order of $\epsilon$,
In order $\epsilon^{\frac{5}{2}}$, from Eqs. \eqref{eq;5}-\eqref{eq:7} and \eqref{eq:14}, we have:
\begin{eqnarray}
&& \frac{\partial n^{(1)}}{\partial \tau}-\gamma_{2}V\frac{\partial n^{(2)}}{\partial\xi}+\gamma_{2}\frac{\partial u^{(2)}}{\partial\xi}+ \gamma_{2}\frac{\partial n^{(1)}u^{(1)}}{\partial\xi}=0 \, ,\label{eq:33}\\
&& \frac{\partial u^{(1)}}{\partial\tau}-\gamma_{2} V \frac{\partial u^{(2)}}{\partial\xi}+\gamma_{2}u^{(1)}\frac{\partial u^{(1)}}{\partial\xi}+ \gamma_{2}\frac{\partial \phi^{(2)}}{\partial\xi}=0 \, ,\label{eq:34}\\
&& \frac{\lambda_{D}^{2}}{L^{2}}\frac{\partial^{2} \phi^{(1)}}{\partial\xi^{2}}=\gamma_{2}(n_{e}^{(2)}-n^{(2)}),\label{eq:35}\\
{\rm and} && \\
&& n_{e}^{(2)}=\int_{-\infty}^{+\infty} f^{(2)}dv \, .\label{eq:36}
\end{eqnarray}
The Vlasov equation in this order reads:
\[-\gamma_{1}\gamma_{2}V\frac{\partial f^{(1)}}{\partial \xi}+v\gamma_{2}^{2}\frac{\partial f^{(2)}}{\partial \xi}+\gamma_{2}^{2}\frac{\partial \phi^{(1)}}{\partial \xi}\frac{\partial f^{(1)}}{\partial v}+\gamma_{2}^{2}\frac{\partial \phi^{(2)}}{\partial \xi}\frac{\partial f^{(0)}}{\partial v}=0 \, ,\]
i.e.
\begin{equation}
v\frac{\partial f^{(2)}}{\partial \xi}+\frac{\partial \phi^{(2)}}{\partial \xi}\frac{\partial f^{(0)}}{\partial v}=\frac{\gamma_{1}}{\gamma_{2}}V\frac{\partial f^{(1)}}{\partial \xi}-\frac{\partial \phi^{(1)}}{\partial \xi}\frac{\partial f^{(1)}}{\partial v} \, .\label{eq:37}
\end{equation}
Now, introducing Eq. \eqref{eq:32a}  into Eq. \eqref{eq:37},
\[v\frac{\partial f^{(2)}}{\partial \xi}+\frac{\partial \phi^{(2)}}{\partial \xi}\frac{\partial f^{(0)}}{\partial v}=-2 \frac{\gamma_{1}}{\gamma_{2}}V\frac{\partial \phi^{(1)}}{\partial \xi}\frac{\partial f^{(0)}}{\partial v^{2}}+2\phi^{(1)}\frac{\partial \phi^{(1)}}{\partial \xi}\frac{\partial}{\partial v}\frac{\partial f^{(0)}}{\partial v^{2}},\]
		i.e.
\[v\frac{\partial f^{(2)}}{\partial \xi}+\frac{\partial \phi^{(2)}}{\partial \xi}\frac{\partial f^{(0)}}{\partial v}=-2 \frac{\gamma_{1}}{\gamma_{2}}V\frac{\partial \phi^{(1)}}{\partial \xi}\frac{\partial f^{(0)}}{\partial v^{2}}+2\phi^{(1)}\frac{\partial \phi^{(1)}}{\partial \xi}\left(\frac{d v^{2}}{d v}\right)\frac{\partial^2 f^{(0)}}{\partial (v^{2})^2},\]
and finally
\begin{equation}
\label{eq:38}
v\frac{\partial f^{(2)}}{\partial \xi}+\frac{\partial \phi^{(2)}}{\partial \xi}\frac{\partial f^{(0)}}{\partial v}=-2 \frac{\gamma_{1}}{\gamma_{2}}V\frac{\partial \phi^{(1)}}{\partial \xi}\frac{\partial f^{(0)}}{\partial v^{2}}+4 v \phi^{(1)}\frac{\partial \phi^{(1)}}{\partial \xi}\frac{\partial^{2} f^{(0)}}{\partial (v^{2})^{2}} \, .
\end{equation}
Now, let us set
\[C_1=\frac{\gamma_{1}}{\gamma_{2}}V\frac{\partial \phi^{(1)}}{\partial \xi}, \qquad {\rm and} \qquad D_1=\phi^{(1)}\frac{\partial \phi^{(1)}}{\partial \xi} \, .\]
We can write Eq. \eqref{eq:38} as
\[v\frac{\partial f^{(2)}}{\partial \xi}+\frac{\partial \phi^{(2)}}{\partial \xi}\frac{\partial f^{(0)}}{\partial v}=-2 C_1\frac{\partial f^{(0)}}{\partial v^{2}}+4 v D_1 \frac{\partial^{2} f^{(0)}}{\partial (v^{2})^{2}} \, .\]

Now, proceeding as above to get a unique solution, we introduce a higher order term
%in $\epsilon^{\frac{9}{2}}$, that is $\gamma_{1}\epsilon^{2}\frac{\partial f^{(2)}}{\partial \tau}$, and adding this in
to the above equation, viz.
\begin{equation}
\label{eq:39}
\gamma_{1}\epsilon^{2}\frac{\partial f^{(2)}_{\epsilon}}{\partial \tau}+v\frac{\partial f^{(2)}_{\epsilon}}{\partial \xi}+\frac{\partial \phi^{(2)}}{\partial \xi}\frac{\partial f^{(0)}_{\epsilon}}{\partial v}=-2 C_1\frac{\partial f^{(0)}_{\epsilon}}{\partial v^{2}}+4 v D_1 \frac{\partial^{2} f^{(0)}_{\epsilon}}{\partial (v^{2})^{2}} \, .
\end{equation}
Taking the Fourier transform of the latter equation, we obtain
\[f^{(2)}_{\epsilon}= \left[\frac{-k \frac{\partial f^{(0)}}{\partial v}}{{k v- \omega \gamma_{1}\epsilon^{2}}}\right]\phi^{(2)}-2i\lim_{\epsilon\rightarrow 0}\left[\frac{C_1\frac{\partial f^{(0)}}{\partial v^{2}}+2vD_1\frac{\partial^{2} f^{(0)}}{\partial (v^{2})^{2}}}{k v- \omega \gamma_{1}\epsilon^{2}}\right] \, . \]

%For a unique solution,
To obtain a unique solutios, we need to find $f^{(2)}$  by letting $\epsilon \rightarrow 0$ in the above equation as \(f^{(2)}=\lim_{\epsilon\rightarrow 0} f_{\epsilon}^{(2)} \, .\)

Taking the  Fourier transform of Eq. \eqref{eq:39}, we find
\[f^{(2)}=\lim_{\epsilon\rightarrow 0} \left[\frac{-k \frac{\partial f^{(0)}}{\partial v}}{{k v- \omega \gamma_{1}\epsilon^{2}}}\right]\phi^{(2)}-2i\lim_{\epsilon\rightarrow 0}\left[\frac{C_1\frac{\partial f^{(0)}}{\partial v^{2}}+2vD_1\frac{\partial^{2} f^{(0)}}{\partial (v^{2})^{2}}}{k v- \omega \gamma_{1}\epsilon^{2}}\right] \, , \]
i.e.
\[f^{(2)}=-k\frac{\partial f^{(0)}}{\partial v}\left[P\left(\frac{1}{k v}\right)+i \pi \delta(k v)\right] \phi^{(2)}-2i\left[C_1\frac{\partial f^{(0)}}{\partial v^{2}}+2vD_1\frac{\partial^{2} f^{(0)}}{\partial (v^{2})^{2}}\right]\left[P\left(\frac{1}{k v}\right)+i \pi \delta(k v)\right] \, ,\]
and
\[f^{(2)}+a_{1}\phi^{(2)}=2i\left[C_1\frac{\partial f^{(0)}}{\partial v^{2}}+2vD_1\frac{\partial^{2} f^{(0)}}{\partial (v^{2})^{2}}\right]\left[P\left(\frac{1}{k v}\right)+i \pi \delta(k v)\right].\]
Multiplying the above equation by $i k$ and integrating over $v$, we have
\begin{equation}
\label{eq:40}
i k [n_{e}^{(2)}-a_{1}\phi^{(2)}]=2C_1\int_{-\infty}^{+\infty}\frac{\partial f^{(0)}}{\partial v^{2}}(i\pi k \delta (k v))dv + 4 D_1\int_{ \infty}^{+\infty}\frac{\partial^{2} f^{(0)}}{\partial (v^{2})^{2}} \, (k v P(k v))\, dv \, .
\end{equation}
		
Substituting with the kappa distribution function\cite{liva}
\[ f^{(0)} = \frac{1}{\sqrt{\pi}\theta}\frac{\Gamma(\kappa+1)}{\kappa^{\frac{3}{2}}\Gamma(\kappa-\frac{1}{2})}\left(1+\frac{v^{2}}{\kappa \theta^{2}}\right)^{-\kappa} \qquad {\rm with } \qquad
\theta=\left( \frac{2\kappa-3}{\kappa}\right)^{\frac{1}{2}}  \, , \]
one finds
\[\int_{-\infty}^{+\infty}\frac{\partial^{2} f^{(0)}}{\partial (v^{2})^{2}} \, dv=\frac{1}{4} \frac{\left(\kappa^{2}-\frac{1}{4}\right)}{\left(\kappa-\frac{3}{2}\right)^{2}} \qquad
{\rm and}	 \qquad	 \int_{-\infty}^{+\infty}k\frac{\partial f^{(0)}}{\partial v^{2}}\delta(kv) dv=\frac{1}{2\sqrt{2\pi}} \frac{\Gamma(\kappa+1)}{(\kappa-\frac{3}{2})^{\frac{3}{2}}\Gamma(\kappa-\frac{1}{2})} \, .\]

We take the inverse-Fourier transform of Eq. \eqref{eq:40} and substituting the values of $C_1$ and $D_1$,
we find
\[\frac{\partial n_{e}^{2}}{\partial \xi}-a_{1}\frac{\partial \phi^{(2)}}{\partial\xi}=a_{3}\phi^{(1)}\frac{\partial \phi^{(1)}}{\partial \xi}+a_{2}F^{-1}\left( i \, {\rm sgn}(k)\right) C_{1}, \]
where
\[F^{-1}\left(i \, {\rm sgn}(k)\right) =-\left( \frac{1}{\pi}\right) P\left( \frac{1}{\xi}\right). \]
Here, we have defined the quantities:
\begin{equation}
\label{eq:41}
   a_{1}=\left(\frac{\kappa-\frac{1}{2}}{\kappa-\frac{3}{2}} \right), \qquad
  a_{2}=\frac{\gamma_{1} V}{\sqrt{2\pi}} \frac{\Gamma(\kappa+1)}{(\kappa-\frac{3} {2})^{\frac{3}{2}}\Gamma(\kappa-\frac{1}{2})} ,
\qquad {\rm and} \qquad
a_{3}= \frac{\left(\kappa^{2}-\frac{1}{4}\right)}{\left(\kappa-\frac{3}{2}\right)^{2}} \, ,
\end{equation}		
while ${\rm sgn}(k) = k/|k|$ is the sign of $k$.

%\[(i k)[n_{e}^{(2)}-a_{1}\phi^{(2)}]=C\sqrt{\frac{\pi}{2}}{i|k|}+D\]
Using the convolution theorem
\begin{equation}\label{A2}
P\left( \frac{1}{\xi}\right) \frac{\partial \phi^{(1)}}{\partial \xi}=P\int_{-\infty}^{+\infty}\left( \frac{\partial \phi^{(1)}}{\partial \xi^{'}}\right) \frac{d \xi^{'}}{\xi-\xi^{'}} \, ,
\end{equation}
we obtain
\begin{equation}
\label{eq:44}
\frac{\partial n_{e}^{2}}{\partial \xi}-a_{1}\frac{\partial \phi^{(2)}}{\partial\xi}+\frac{a_{2}}{\gamma_{2}}P\int_{-\infty}^{+\infty}\left( \frac{\partial \phi^{(1)}}{\partial \xi^{'}}\right) \frac{d \xi^{'}}{\xi-\xi^{'}}-a_{3}\phi^{(1)}\frac{\partial \phi^{(1)}}{\partial \xi}=0 \, .
\end{equation}
		
Multiplying Eq. \eqref{eq:33} by $V$ and adding to  Eq. \eqref{eq:34},
\begin{equation}
\label{eq:45}
\frac{\partial u^{(1)}}{\partial\tau}+V\frac{\partial n^{(1)}}{\partial\tau}+\gamma_{2}u^{(1)}\frac{\partial u^{(1)}}{\partial \xi}+\gamma_{2}\frac{\partial \phi^{(2)}}{\partial \xi}-\gamma_{2}V^{2}\frac{\partial n^{(2)}}{\partial \xi}+\gamma_{2}V\frac{\partial n^{(1)}u^{(1)}}{\partial \xi}=0 \, .
\end{equation}
		
Differentiating both sides of Eq. \eqref{eq:35} (i.e. operating by $\frac{\partial}{\partial \xi}$),
\[\gamma_{3}\frac{\partial^{3} \phi^{(1)}}{\partial\xi^{3}}=\gamma_{2}\left( \frac{\partial n_{e}^{(2)}}{\partial \xi}-\frac{\partial n^{(2)}}{\partial \xi}\right) ,\]
i.e.
\begin{equation}
\label{eq:46}
\gamma_{2}\frac{\partial n^{(2)}}{\partial \xi}=\gamma_{2}\frac{n_{e}^{(2)}}{\partial \xi}-\gamma_{3}\frac{\partial^{3} \phi^{(1)}}{\partial\xi^{3}} \, .
\end{equation}
Now, combining Eq. \eqref{eq:46} into Eq. \eqref{eq:45},
\[\frac{\partial u^{(1)}}{\partial\tau}+V\frac{\partial n^{(1)}}{\partial\tau}+\gamma_{2}u^{(1)}\frac{\partial u^{(1)}}{\partial \xi}+\gamma_{2}\frac{\partial \phi^{(2)}}{\partial \xi}-V^{2}\gamma_{2}\frac{n_{e}^{(2)}}{\partial \xi}+\gamma_{3}V^{2}\frac{\partial^{3} \phi^{(1)}}{\partial\xi^{3}}+\gamma_{2}V\frac{\partial n^{(1)}u^{(1)}}{\partial \xi}=0 \, .\]
Substituting $\frac{\partial n_{e}^{2}}{\partial \xi}$ from Eq. \eqref{eq:44},
\begin{eqnarray*}
\frac{\partial u^{(1)}}{\partial\tau}+V\frac{\partial n^{(1)}}{\partial\tau}+\gamma_{2}u^{(1)}\frac{\partial u^{(1)}}{\partial \xi}+\gamma_{2}\frac{\partial \phi^{(2)}}{\partial \xi}-V^{2}\gamma_{2}\left[a_{1}\frac{\partial \phi^{(2)}}{\partial\xi}-\frac{a_{2}}{\gamma_{2}}P\int_{-\infty}^{+\infty}\left( \frac{\partial \phi^{(1)}}{\partial \xi^{'}}\right) \frac{\partial \xi^{'}}{\xi-\xi^{'}}+a_{3}\phi^{(1)}\frac{\partial \phi^{(1)}}{\partial \xi} \right]\\
+\gamma_{3}V^{2}\frac{\partial^{3} \phi^{(1)}}{\partial\xi^{3}}+\gamma_{2}V\frac{\partial n^{(1)}u^{(1)}}{\partial \xi}=0 .
\end{eqnarray*}
		%
%To obtain the required KdV equation, we
One may now eliminate $\frac{\partial u^{(2)}}{\partial \xi}$ and $\frac{\partial n^{(2)}}{\partial \xi}$ by using Eqs. \eqref{eq:33}- \eqref{eq:35} and then substituting the values of $\phi^{(1)}$ and $u^{(1)}$ from Eqs. \eqref{eq:27}:
		\[\phi^{(1)}=V^{2}n^{(1)} \qquad {\rm and} \qquad u^{(1)}=Vn^{(1)} \, .\]
One thus obtains
\[2 V \frac{\partial n^{(1)}}{\partial\tau}+\gamma_{2}V^{2}n^{(1)}\frac{\partial n^{(1)}}{\partial \xi}+{a_{2}}V^{4}P\int_{-\infty}^{+\infty}\left( \frac{\partial n^{(1)}}{ \partial \xi^{'}}\right) \frac{d \xi^{'}}{\xi-\xi^{'}}-a_{3}V^{6}\gamma_{2}n^{(1)}\frac{\partial n^{(1)}}{\partial \xi}
+\gamma_{3}V^{4}\frac{\partial^{3} n^{(1)}}{\partial\xi^{3}}+2\gamma_{2}V^{2}n^{(1)}\frac{\partial n^{(1)}}{\partial \xi}=0 \, ,\] 		
and
\[2 V \frac{\partial n^{(1)}}{\partial\tau}+3\gamma_{2}V^{2}n^{(1)}\frac{\partial n^{(1)}}{\partial \xi}+{a_{2}}V^{4}P\int_{-\infty}^{+\infty}\left( \frac{\partial n^{(1)}}{\partial \xi^{'}}\right) \frac{d\xi^{'}}{\xi-\xi^{'}}-a_{3}V^{6}\gamma_{2}n^{(1)}\frac{\partial n^{(1)}}{\partial \xi} +\gamma_{3}V^{4}\frac{\partial^{3} n^{(1)}}{\partial\xi^{3}}=0 \, ,\]
		or
\[\frac{\partial n^{(1)}}{\partial\tau}+\frac{\gamma_{2}}{2V}\left[3V^{2}-a_{3}V^{6} \right] n^{(1)}\frac{\partial n^{(1)}}{\partial \xi}+\frac{{a_{2}}V^{4}}{2V}P\int_{-\infty}^{+\infty}\left( \frac{\partial n^{(1)}}{ \partial\xi^{'}}\right) \frac{d \xi^{'}}{\xi-\xi^{'}}
	+\frac{\gamma_{3}V^{3}}{2}\frac{\partial^{3} n^{(1)}}{\partial\xi^{3}}=0 \, . \]

The latter expression is a modified KdV equation (with an extra term in account of Landau damping) in the form:
\begin{equation}
  \label{eq:47}
\frac{\partial n}{\partial\tau}+A n\frac{\partial n}{\partial \xi}+B\frac{\partial^{3} n}{\partial\xi^{3}}+C P\int_{-\infty}^{+\infty}\left( \frac{\partial n} {\partial \xi^{'}}\right) \frac{d\xi^{'}}{\xi-\xi^{'}}=0 \, ,
\end{equation}
i.e. precisely Eq. (\ref{eq:KdV}).

\end{document}